\documentclass[letterpaper,10pt]{article}

\usepackage{style}
\usepackage[english]{babel} 
\usepackage{amsmath,amssymb,amsfonts}
\usepackage{hyperref}
\usepackage{dsfont}

\numberwithin{equation}{section}
\DeclareMathAlphabet{\mathcal}{OMS}{cmsy}{m}{n}

\begin{document}

\title{
A Jones matrix formalism for simulating three-dimensional polarized light imaging \\of brain tissue}

\author{
M. Menzel$^{1,}$*, K. Michielsen$^{2}$, H. De Raedt$^{3}$, J. Reckfort$^{1}$, K. Amunts$^{1,4}$ and M. Axer$^{1}$}

\address{
$^1$Institute of Neuroscience and Medicine (INM-1) and $^2$J\"ulich Supercomputing Centre,\\
Forschungszentrum J\"ulich, Wilhelm-Johnen-Stra{\ss}e, J\"ulich 52425, Germany\\
$^3$ Department of Applied Physics, Zernike Institute for Advanced Materials, University of Groningen, Nijenborgh 4, Groningen 9747 AG, The Netherlands\\
$^4$ C{\'e}cile and Oskar Vogt Institute of Brain Research, University of D\"usseldorf, D\"usseldorf 40204, Germany}

\email{
* author for correspondence (m.menzel@fz-juelich.de)}


\begin{abstract}

The neuroimaging technique three-dimensional polarized light imaging (3D-PLI) provides a high-resolution reconstruction of nerve fibres in human post-mortem brains. The orientations of the fibres are derived from birefringence measurements of histological brain sections assuming that the nerve fibres -- consisting of an axon and a surrounding myelin sheath -- are uniaxial birefringent and that the measured optic axis is oriented in direction of the nerve fibres (macroscopic model). Although experimental studies support this assumption, the molecular structure of the myelin sheath suggests that the birefringence of a nerve fibre can be described more precisely by multiple optic axes oriented radially around the fibre axis (microscopic model).

In this paper, we compare the use of the macroscopic and the microscopic model for simulating 3D-PLI by means of the Jones matrix formalism.
The simulations show that the macroscopic model ensures a reliable estimation of the fibre orientations as long as the polarimeter does not resolve structures smaller than the diameter of single fibres. In the case of fibre bundles, polarimeters with even higher resolutions can be used without losing reliability. When taking the myelin density into account, the derived fibre orientations are considerably improved.

\vspace{0.5mm}

\center{\textbf{Keywords: polarized light imaging; nerve fibre architecture; optics; birefringence; \\ Jones matrix calculus; computer simulation}}
\end{abstract}


\section{Introduction}
\label{sec:intro}

Unravelling the architecture and connectivity of nerve fibres in the human brain is one of the greatest challenges in neuroscience. Over the past years, several methods have been developed to reconstruct the human connectome \cite{behrens12, sporns05, sporns}. The neuroimaging technique \textit{three-dimensional polarized light imaging (3D-PLI)} has been employed to reconstruct the three-dimensional architecture of nerve fibres in human post-mortem brains with a resolution of a few micrometres \cite{axer11_1, axer11_2}. 3D-PLI enables the investigation of the pathways of long-range fibre bundles as well as single fibres and thus serves as a bridging technology between the macroscopic and the microscopic scale. 

The spatial orientations of the nerve fibres are derived by transmitting polarized light through histological brain sections in a polarimeter and measuring their birefringence. 
To relate the measured signal to the fibre orientation, an \textit{effective model} of birefringence is used which assumes that the fibre density is constant over the whole brain section \cite{axer11_1} and that the measured optic axis indicates the predominant fibre orientation \cite{axer11_2,dohmen15}. This assumption is based on various experimental studies on white matter which show that the average birefringence of parallel nerve fibres is negatively uniaxial and that the measured optic axis is oriented along the length of the fibres \cite{goethlin13, schmitt39, vidal80, ambronn90}. 

The majority of nerve fibres in the brain consist of an axon and a surrounding myelin sheath.
The cytoplasm of the axon contains tubular polymers (microtubules) and neurofilaments running along the length of the axon \cite{darnell,yuan12}.
The myelin sheath is formed by oligodendrocytes (glial cells) which are spirally wrapped around the axon. The cell membranes are bimolecular layers consisting of lipid molecules and membrane proteins. The membrane proteins are embedded in the bilayer or attached to the membrane surface \cite{quarles06,aggarwal11,martenson}, whereas the lipid molecules are oriented radially to the fibre axis \cite{martenson,morell,bear71}. 
The cell organelles of the axon and the protein framework of the myelin sheath lead to a weak positive birefringence with respect to the longitudinal fibre axis \cite{goethlin13,schmitt39,bear36,huang05,oldenbourg98,ambronn90,bear71}. The anisotropic structure of the lipid molecules causes a positive birefringence with respect to the radial fibre axis \cite{goethlin13, bear36, schmitt39, martenson, koike-tani13}.

The effective model of uniaxial negative birefringence that is currently used in 3D-PLI seems reasonable for sufficiently low optical resolutions. However, it might no longer be valid if the anisotropic molecular structure of the nerve fibres is resolved. In this paper, we investigated the limitations of the effective model in terms of the optical resolution of the polarimeter using numerical simulations. 
The simulations were performed with a modified version of SimPLI \cite{dohmen15}, a simulation method that models the birefringence of the fibres with the Jones matrix calculus and allows data to be generated from synthetic fibre constellations that is comparable to experimental data. 
In order to study and understand the most dominant effects that generate the birefringence signals in 3D-PLI, the anisotropic molecular structure of the nerve fibres was described by a simplified birefringence model with radial optic axes (\textit{microscopic model}) and the effective model of uniaxial negative birefringence by a birefringence model with axial optic axes (\textit{macroscopic model}). To investigate the limitations of the effective model, the transition between the microscopic and the macroscopic model was investigated depending on the optical resolution of the imaging system.


\section{Three-dimensional polarized light imaging (3D-PLI)}

The neuroimaging technique 3D-PLI determines the orientation of nerve fibres in post-mortem brains at the micrometre scale. The principles of 3D-PLI have been explained in detail by \textsc{Larsen} \textit{et al.} \cite{larsen07} and \textsc{Axer} \textit{et al.} \cite{axer11_1,axer11_2}. This section describes the measurement and data analysis procedures that are relevant for this study.


\subsection{Measurement}

To determine the orientation of the nerve fibres, a post-mortem brain --  obtained from a body donor in accordance with ethical requirements -- is fixed in buffered formaldehyde for several months, frozen and cut with a cryotome into histological sections of 70\,\textmu m, which are measured with a polarimeter.
For the 3D-PLI measurement, two state-of-the-art polarimeters with different optical resolutions and sensitivities are employed: The \textit{large-area polarimeter (LAP)} has a pixel size of $64$\,\textmu m and is mainly used for single-shot images of whole human brain sections. The \textit{polarizing microscope (PM)} has a pixel size of $1.33$\,\textmu m (i.\,e.\ down to small axonal diameters), which enables complex fibre constellations to be disentangled.

The LAP contains a pair of crossed linear polarizers and a quarter-wave retarder (with its fast axis adjusted at an angle of $-45^{\circ}$ with respect to the transmission axis of the first linear polarizer), see Fig.\ \ref{fig:3DPLI}a. The employed light source emits incoherent, non-polarized, diffusive light with a peak wavelength of 525\,nm.
During the measurement, the polarizers and the quarter-wave retarder are rotated simultaneously around the stationary tissue sample. For each rotation angle $\rho = 0^{\circ}, 10^{\circ}, ..., 170^{\circ}$, the transmitted light intensity is recorded by a CCD camera so that a series of 18 images is acquired.

The imaging principle works as follows: The quarter-wave retarder transforms the linearly polarized light from the first polarizer into circularly polarized light. The birefringent brain tissue induces an additional phase shift so that the outgoing light is elliptically polarized. The fraction of light that then passes the second linear polarizer depends on the local orientation of the optic axis of the birefringent tissue, which is assumed to coincide with the local fibre orientation. 

The polarimetric set-up of the PM is slightly different to the set-up of the LAP (the order of the optical elements is reversed and only the first linear polarizer is rotatable). However, the imaging principle and the signal analysis are similar \cite{axer11_1} so that the following considerations are only described for the LAP.


\subsection{Signal analysis}
\label{sec:3DPLI_analysis}

\begin{figure}[htbp]
\centering
\includegraphics[width=1 \textwidth]{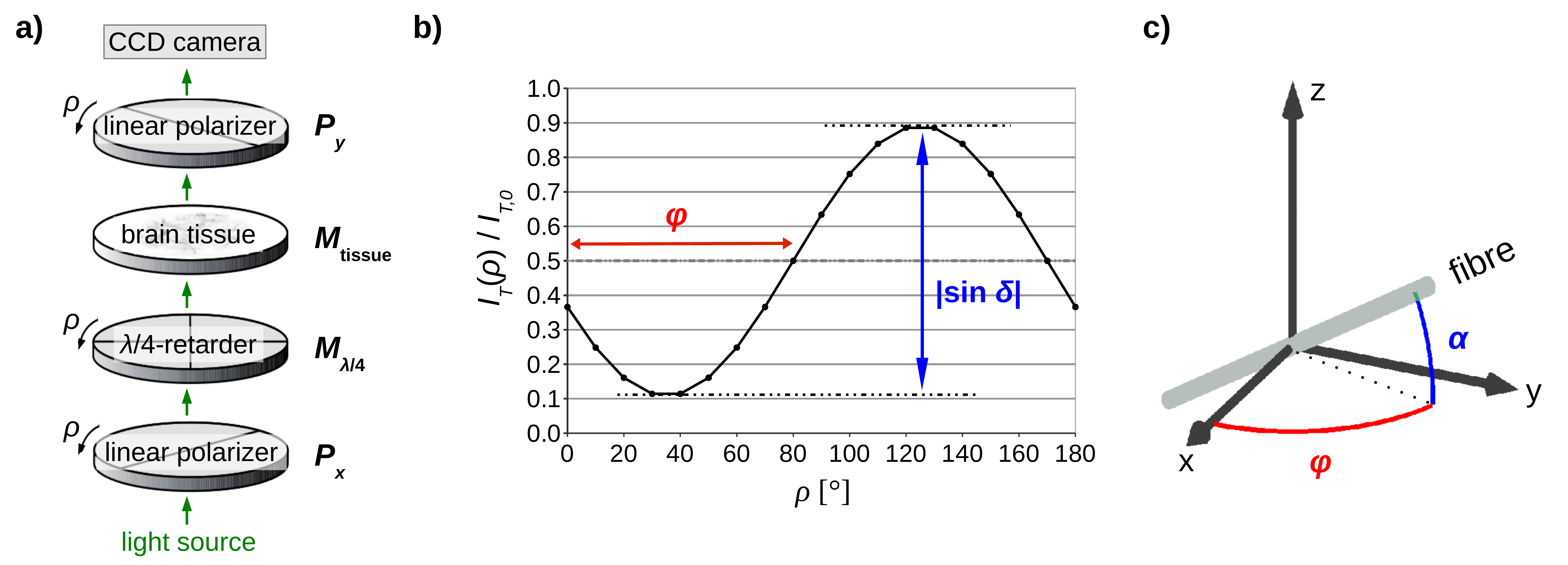}
\caption{ \textbf{(a)} Measurement set-up of 3D-PLI (for the LAP): The brain tissue is placed between a pair of crossed linear polarizers and a quarter-wave retarder, which are rotated simultaneously by 18 discrete rotation angles $\rho$. The transmitted light intensity is calculated with the Jones calculus, in which each optical element is represented by a Jones matrix (bold symbols). \textbf{(b)} The normalized transmitted light intensity $I_T(\rho) / I_{T,0}$ describes a sinusoidal curve for each image pixel. The phase $\varphi$ corresponds to the local fibre direction angle and the peak-to-peak amplitude $\lvert \sin \delta \rvert$ to the local fibre inclination angle. \textbf{(c)} The three-dimensional orientation of a fibre is defined by the direction angle $\varphi$ and the inclination angle $\alpha$.}
\label{fig:3DPLI}
\end{figure}

The measured light intensity of an individual pixel describes a sinusoidal curve across the acquired image series, which depends on the orientation of the fibres within this pixel (see Fig.\ \ref{fig:3DPLI}b). 
A physical description of the measured light intensity profile can be derived with the Jones matrix calculus \cite{jones41,jones42}, assuming that the light is coherent and completely polarized and that the optical elements are linear.
For simplicity, the derivation is shown for a single pixel at a certain rotation angle $\rho$.

In the Jones matrix calculus, all optical elements in the polarimeter are represented by Jones matrices (cf.\ Fig.\ \ref{fig:3DPLI}a).
The Jones matrices of the crossed linear polarizers are given by \cite{collett}:
\begin{align}
P_x =
	\begin{pmatrix} 1 	&  0  \\ 
                	0 	&  0
    \end{pmatrix}
\,\, , \,\,\,\,\,\,\,
P_y = 
	\begin{pmatrix} 0 	&  0  \\ 
                	0 	&  1
    \end{pmatrix} .
\label{eq:polarizers}
\end{align}
The Jones matrix of a wave retarder that is rotated by an angle $\psi$ in counterclockwise direction and induces along the fast axis a phase shift $\delta$ between the two orthogonal components of the light wave is given by \cite{collett}:
\begin{align}
M_{\delta}(\psi) &= R(\psi) \cdot M_{\delta} \cdot R(-\psi) \notag \\
&=	\begin{pmatrix} \cos\psi  & -\sin\psi \\ 
                	\sin\psi  & \cos\psi
    \end{pmatrix} \,
	\begin{pmatrix} e^{\operatorname{i} \delta/2} &  0 			\\ 
                	0 			 &  e^{-\operatorname{i} \delta/2}
    \end{pmatrix}
    \begin{pmatrix} \cos\psi  & \sin\psi \\ 
                	-\sin\psi & \cos\psi
    \end{pmatrix} .
\label{eq:M_retarder} 
\end{align}
In the experimental set-up, the fast axis of the quarter-wave retarder is rotated by $-45^{\circ}$ with respect to the axis of the first linear polarizer. Thus, the quarter-wave retarder can be described by the Jones matrix of a rotated wave retarder as given in Eq.\ (\ref{eq:M_retarder}) with a rotation angle of $\psi = -45^{\circ}$ and a phase shift of $\delta = 90^{\circ}$:
\begin{align}
M_{\lambda/4} 
\equiv M_{90^{\circ}}(-45^{\circ})
= \frac{1}{\sqrt{2}}\,
\begin{pmatrix}
1 & -\operatorname{i} \\
-\operatorname{i} & 1
\end{pmatrix}.
\label{eq:M_lambda/4}
\end{align}

Under the assumption that the birefringence of the brain tissue can locally be described as negatively uniaxial with the optic axis indicating the predominant fibre direction (effective model), the brain tissue can locally be represented by a wave retarder that  introduces a phase shift $\delta$ along the fast axis (fibre axis). 
During the measurement, the two polarizers and the quarter-wave retarder are rotated simultaneously around the specimen stage in counterclockwise direction by a rotation angle $\rho$. For simplicity, the equivalent case is considered in which the brain tissue is rotated by an angle $(-\rho )$ in counterclockwise direction while the other optical elements are fixed. Thus, the brain tissue can be described by the Jones matrix of a rotated wave retarder as given in Eq.\ (\ref{eq:M_retarder}) with phase shift $\delta$ and rotation angle $ \psi = \varphi - \rho$, where $\varphi$ denotes the in-plane orientation of the optic axis:
\begin{align}
M_{\text{tissue}} 
\equiv M_{\delta}(\varphi - \rho)
= \begin{pmatrix} \cos(\varphi - \rho)  & -\sin(\varphi - \rho) \\ 
                	\sin(\varphi - \rho)  & \cos(\varphi - \rho)
    \end{pmatrix} \,
	\begin{pmatrix} e^{\operatorname{i} \delta/2} &  0 			\\ 
                	0 			 &  e^{-\operatorname{i} \delta/2}
    \end{pmatrix}
    \begin{pmatrix} \cos(\varphi - \rho)  & \sin(\varphi - \rho) \\ 
                	-\sin(\varphi - \rho) & \cos(\varphi - \rho)
    \end{pmatrix} .
\label{eq:M_tissue} 
\end{align}

When light with an electric field vector $\vec{E}_0$ passes through the 3D-PLI set-up, the resulting output beam with electric field vector $\vec{E_T}$ can be described by multiplication of the associated Jones matrices.
As the Jones matrix calculus cannot be used to describe the non-polarized light emitted by the employed light source, the Jones vector $\vec{E}_x = P_x \cdot \vec{E}_0$ is used to describe the horizontally polarized light after the first linear polarizer (cf.\ Fig.\ \ref{fig:3DPLI}a):
\begin{align}
\vec{E_T} = P_y \cdot M_{\text{tissue}} \cdot M_{\lambda /4} \cdot \vec{E}_x.
\label{eq:E}
\end{align}
Using $I_T \sim \vert\vec{E_T}\vert^2$, the transmitted light intensity is calculated, yielding a sinusoidal intensity profile (see Fig.\ \ref{fig:3DPLI}b):
\begin{align}
I_T(\rho) &= \frac{I_{T,0}}{2}\,\Big(1 + \sin\big(2(\rho - \varphi)\big)\,\sin\delta \Big),
\label{eq:intensity}
\end{align}
where $I_{T,0} \sim \vert\vec{E_x}\vert^2$ corresponds to the transmitted light intensity averaged over all rotation angles (here referred to as \textit{transmittance}) and $\vert\sin\delta\vert$ to the peak-to-peak amplitude of the normalized sinusoidal intensity profile (here referred to as \textit{retardation}).
The phase shift $\boldsymbol{\delta}$ is given by (see Appx.\ \ref{appx:derivation_PhaseShift}):
\begin{align}
\delta &\approx \frac{2\pi}{\lambda}\, t \, \Delta n \, \cos ^2\alpha \,,
\label{eq:phaseshift}
\end{align}
where $\lambda$ is the wavelength of the incident light, $t$ the thickness of the brain section, $\Delta n$ the local birefringence of the sample and $\boldsymbol{\alpha}$ the local out-of-plane inclination angle of the fibre. 
Thus, the intensity profile in Eq.\ (\ref{eq:intensity}) is a direct measure of the spatial fibre orientation defined by the direction angle $\varphi$ and the inclination angle $\alpha$ (see Fig.\ \ref{fig:3DPLI}c).

In order to compute transmittance, direction and retardation, the intensity profile is fitted by means of a discrete harmonic Fourier analysis \cite{axer11_1, glazer96}. The inclination angle $\alpha$ is calculated from the measured retardation $\lvert \sin \delta \rvert$ by rearranging Eq.\ (\ref{eq:phaseshift}).
The direction and inclination angles are combined to a unit vector indicating the local fibre orientation in three dimensions. 
Putting all unit vectors of several adjacent brain sections together, a three-dimensional volume of vectors is created and the fibre tracts are reconstructed with streamline algorithms.


\section{Simulation of 3D-PLI using the Jones matrix formalism}

\subsection{Simulation model}
\label{sec:simulation_models}

3D-PLI derives the nerve fibre orientations based on the fact that the average birefringence of parallel fibres is negatively uniaxial \cite{goethlin13, schmitt39, vidal80, ambronn90} and assuming that the orientation of the measured optic axis corresponds to the local fibre orientation.
To investigate the limitations of this effective birefringence model, a straight single fibre and a hexagonal bundle of straight parallel fibres were simulated and the birefringence of the fibres was modelled according to a microscopic and a macroscopic model for different optical resolutions of the simulated imaging system. 
\paragraph{\textit{Microscopic model:}}
The microscopic model of birefringence considers the anisotropic molecular structure of a single nerve fibre. To investigate and understand the predominant effects generating the birefringence signals in 3D-PLI, a simplified model of birefringence was chosen for the simulations. As stated in Sec.\ \ref{sec:intro}, the average birefringence of parallel nerve fibres is negative with respect to the longitudinal fibre axis. Therefore, the positive birefringence of the axon and the myelin proteins is weak as compared to the birefringence effects of the myelin lipids \cite{bear36,schmitt39,vidal80,oldenbourg98,koike-tani13}. Since the exact contribution of the different birefringence effects to the overall birefringence is unknown, the birefringence effects of the nerve fibres were modelled by considering only the anisotropic radial structure of the myelin sheaths: The fibres were simulated as hollow tubes (representing the myelin sheaths) with positive birefringence and radial optic axes (cf.\ lower Fig.\ \ref{fig:simulation_method}b). The axons were considered to be non-birefringent.
\paragraph{\textit{Macroscopic model:}}
To compare the simulation results of the microscopic model with the effective model of uniaxial negative birefringence, a macroscopic model of birefringence was defined. According to the assumptions made in the effective model, a single nerve fibre was simulated as negatively birefringent with axial optic axes oriented along the length of the fibre (cf.\ upper Fig.\ \ref{fig:simulation_method}b). 
\textsc{Dohmen} \textit{et al.} \cite{dohmen15} used this simulation model to investigate the effect of crossing fibre constellations. As this study concentrates on straight parallel fibres, the macroscopic model only serves as a reference for the effective model to verify the simulations of the microscopic model. To ensure a better comparison with the microscopic model, the fibres were simulated as hollow tubes (and not as solid cylinders as in \cite{dohmen15}).


\subsection{Simulation method}

The basic idea of the simulation method is to model the birefringent myelin sheaths as series of linear optical retarder elements which are represented by Jones matrices. By defining the direction of the optic axes (radial/axial), both the microscopic and the macroscopic model can be simulated. 
The simulation approach is based on the simulation tool SimPLI developed by \textsc{Dohmen} \textit{et al.} \cite{dohmen15}. 
For this study, the simulation tool was extended by the microscopic model and modified such that various fibre configurations with individual orientations, radii and myelin sheath thickness can be realized. 

The simulation tool is based on several assumptions and simplifications: 
First of all, the use of the Jones matrix calculus requires linear optical elements and perfect polarizers (i.\,e.\ the outgoing light is assumed to be completely polarized). Another assumption is that the incident light can be described by parallel rays of light with straight optical pathways, i.\,e.\ the light is assumed to be non-diffusive and refraction, diffraction and scattering are neglected. 
For this study, a parallel and straight beam of light seems a reasonable approximation for the LAP because the imaging system has a small numerical aperture (the acceptance angle of the objective lens is less than $1^{\circ}$) so that the camera only captures light rays that are almost parallel to each other.

\begin{figure}[htbp]
\centering
\includegraphics[width=1 \textwidth]{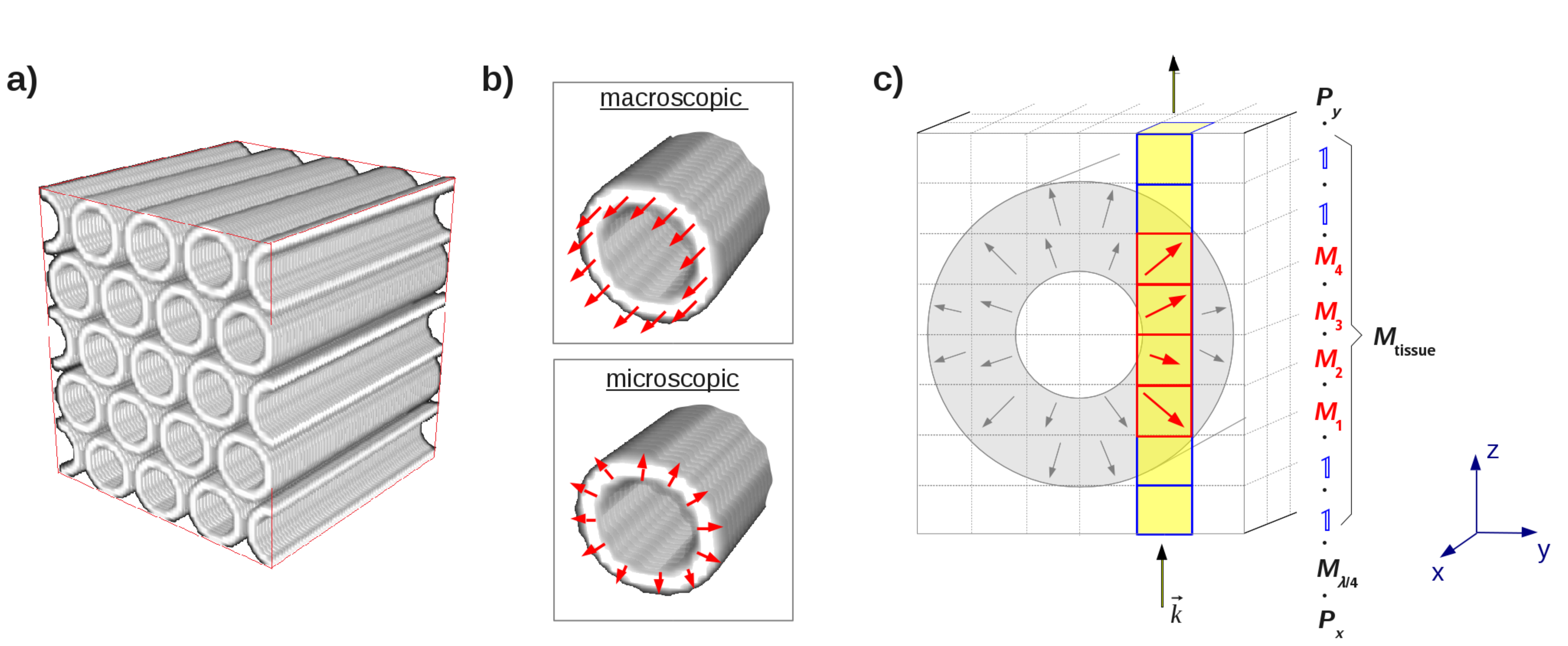}
\caption{Simulation method: \textbf{(a)} Generation of synthetic nerve fibres in a three-dimensional volume. \textbf{(b)} Generation of a three-dimensional vector field according to the macroscopic model (axial optic axes) and the microscopic model (radial optic axes). \textbf{(c)} Generation of a synthetic 3D-PLI image series (illustrated for a large fibre in the microscopic model): The simulation volume is discretized into small volume elements (voxels). Each myelin voxel (grey) is represented by the Jones matrix of an optical retarder ($M_j$) whose axis is oriented in direction of the optic axes (arrows). The polarizing filters of the 3D-PLI set-up (see Fig.\ \ref{fig:3DPLI}a) are also represented by Jones matrices. For each rotation angle of the polarimeter, all Jones matrices along the optical path (highlighted column) are multiplied.}
\label{fig:simulation_method}
\end{figure}

The simulation consists of several steps:
\paragraph{\textit{1.) Generation of synthetic nerve fibres in a three-dimensional volume:}} The nerve fibres are modelled as hollow tubes representing the myelin sheaths (see Fig.\ \ref{fig:simulation_method}a).
In order to approximate the geometry of the fibres, the simulation volume is discretized into small cubic volume elements (called \textit{voxels}), as indicated schematically by the grid in Fig.\ \ref{fig:simulation_method}c.
\paragraph{\textit{2.) Generation of a three-dimensional vector field:}} For sufficiently small voxel sizes, the birefringence of the myelin sheaths can approximately be described by assigning each myelin voxel $j$ a unit vector that indicates the direction of the optic axis $(\varphi_j, \alpha_j)$ within the myelin sheath. In the macroscopic model, the vectors are oriented parallel to the fibre axis. In the microscopic model, the vectors are oriented radially to the fibre axis (see Fig.\ \ref{fig:simulation_method}b).
\paragraph{\textit{3.) Generation of a synthetic 3D-PLI image series:}}
In order to model the birefringence effect of the myelin sheaths, each myelin voxel is represented by the Jones matrix of a rotated wave retarder. The retarder axis is aligned with the optic axis within the myelin voxel (see Fig.\ \ref{fig:simulation_method}c).
The synthetic 3D-PLI image series is calculated analogously to the derivation of the sinusoidal intensity profile as given in Eq.\ (\ref{eq:E}), with $M_{\text{tissue}}$ being replaced by the product of $N$ matrices representing the myelin voxels along the optical path:
\begin{align}
\vec{E_T} = P_y \cdot (M_{N} \cdot M_{N-1} \cdots M_1) \cdot M_{\lambda /4} \cdot \vec{E}_x.
\label{eq:E_sim}
\end{align}
The matrix $M_j \equiv M_{\delta_j}(\varphi_j - \rho)$ is the Jones matrix of a rotated wave retarder as given in Eq.\ (\ref{eq:M_retarder})  and represents the $j$-th myelin voxel. The rotation angle depends on the in-plane direction angle $\varphi_j$ of the optic axis and the phase shift $\delta_j$ on the out-of-plane inclination angle $\alpha_j$.
The Jones matrices of the linear polarizers and the quarter-wave retarder are given by Eqs.\ (\ref{eq:polarizers}) and (\ref{eq:M_lambda/4}). For each rotation angle of the polarimeter ($\rho = 0^{\circ}, 10^{\circ}, ..., 170^{\circ}$), all Jones matrices along the optical path are multiplied (see Fig.\ \ref{fig:simulation_method}c), yielding a series of 18 synthetic 3D-PLI images with a sinusoidal intensity profile for each image pixel. 

\subsection{Simulation parameters}
\label{sec:simulation_parameters}

The choice of the simulation parameters was inspired by real experimental conditions. According to typical dimensions of large nerve fibres in human white matter \cite{aboitiz92,longstaff00,morell,hildebrand92}, the diameter and the myelin sheath thickness of the simulated nerve fibres were chosen to be $15$\,\textmu m and $2.5$\,\textmu m, respectively  (see Fig.\ \ref{fig:simulated_fibre}a).
The fibres were generated in a simulation volume with dimensions $x \times y \times z = 64 \times 64 \times 70$\,\textmu m$^3$, corresponding to the pixel size of the LAP and the thickness of the brain section. The simulation volume was discretized into cubic voxels with a side length of $\Delta x_{\text{sim}}$. In a preliminary study (see later, Sec.\ \ref{sec:results_comparison}), the optimal voxel size was determined to be $\Delta x_{\text{sim}} = 0.1$\,\textmu m, which was used for all following simulations.
Note that the dimensions are given in micrometres to meet the experimental conditions. As only relative length scales matter for the qualitative simulation results, the units could be chosen arbitrarily.

Since measuring the birefringence of the micrometre-thick brain sections is impossible with the employed set-ups and literature values are not given for the currently used preparation technique, an upper limit for the birefringence of the myelin sheaths $\Delta n$ was estimated: Under the assumption that a brain section that is completely filled with a homogeneous birefringent material with in-plane optic axis ($\alpha=0^{\circ}$) induces a maximum possible retardation ($\vert\sin\delta\vert = 1 \Leftrightarrow \delta = \pi / 2$), the upper limit of the birefringence was calculated by rearranging Eq.\ (\ref{eq:phaseshift}): $\Delta n = \lambda / (4t) = (525$\,nm$) / (4 \cdot 70$\,\textmu m$) \approx 0.001875$. Note that the choice of $\Delta n$ only changes the overall magnitude of the retardation and does not affect the simulation results qualitatively. In the macroscopic (microscopic) model, the myelin voxels were simulated with axial (radial) optic axes and negative (positive) birefringence with respect to the optic axes.

The wavelength of the incident light was chosen to correspond to the peak wavelength of the LAP ($\lambda = 525$\,nm).
To study only the birefringence effect of the nerve fibres, the fibres were simulated without any absorption.


\subsection{Simulation of the optical resolution}
\label{sec:simulation_optical_resolution}

To investigate the effect of different optical resolutions on the measured 3D-PLI signal, the synthetic 3D-PLI image series were downsampled using the open-source image processing programme Fiji \cite{fiji}: To account for the limited optical resolution of the polarimeter, the image series were first convoluted with a two-dimensional Gaussian filter with a standard deviation $\sigma$. Then, the effect of the spatial discretisation of the CCD chip was modelled by resampling the resulting images with a sampling factor $f_{\text{s}}$ (average when downsizing). To determine realistic parameters for $\sigma$ and $f_{\text{s}}$, the imaging properties of the LAP were considered as a point of reference (see Appx.\ \ref{appx:OpticalResolution}).

Based on these considerations, the synthetic 3D-PLI image series were downsampled with different parameter sets (see Tab.\ \ref{tab:downsampling}), yielding images with different pixel sizes $\Delta x$. The pixel size of the downsampled images was chosen such that a multiple of the pixel size corresponds to the side length of the simulation volume ($\Delta x = 64$\,\textmu m$/n$, with $n = 4,\,8,\,16,\,32$). The standard deviation was calculated as a linear function of the pixel size ($\sigma = 0.714\,\Delta x$, see Appx.\ \ref{appx:OpticalResolution}) and the sampling factor was calculated by dividing the pixel size of the high-resolution image series by the pixel size of the downsampled image ($f_{\text{s}} = \Delta x_{\text{sim}} / \Delta x = 0.1$\,\textmu m $/\Delta x$). In the following, the optical resolution of the imaging system will be given in terms of the pixel size, which defines the set of downsampling parameters ($\sigma$ and $f_{\text{s}}$) in Tab.\ \ref{tab:downsampling}.
Note that the simulation results will not change qualitatively as long as the ratio between the fibre dimensions and the downsampling parameters remains the same.

 \begin{table}[htbp] 
 \centering 
 \begin{tabular}{|r|r|l|} 
 \hline 
 $\Delta x$ [\textmu m] & $\sigma$ [\textmu m]  & \,\,\,$f_{\text{s}}$ \,\\ 
 \hline 
 \hline 
 2.00\,\,\,	 	& 1.43\,\, 	& 1/20	\\
 4.00\,\,\,	 	& 2.86\,\, 	& 1/40	\\
 8.00\,\,\,	 	& 5.71\,\, 	& 1/80	\\
 16.00\,\,\, 	& 11.43\,\,	& 1/160	\\
\hline 
 \multicolumn{3}{c}{} 
 \end{tabular} 
 \caption{Downsampling parameters (selected values): To obtain an image with pixel size $\Delta x$, a two-dimensional Gaussian filter with standard deviation $\sigma = 0.714\,\Delta x$ and resampling with sampling factor $f_{\text{s}} = 0.1\,{\text{\textmu m}}/\Delta x$ are applied to the image.} 
 \label{tab:downsampling} 
 \end{table} 


\subsection{Calculation of the retardation curve}
\label{sec:simulation_RetardationCurve}

The determination of the inclination angle $\alpha$ is challenging for 3D-PLI because the peak-to-peak amplitude of the measured intensity profile ($\vert\sin\delta\vert$) is highly sensitive to noise and -- amongst others -- influenced by the density of myelinated nerve fibres (see below).

In the standard 3D-PLI analysis, the inclination angle is calculated from the measured intensity profile assuming that the brain tissue can locally be described by the effective model of uniaxial negative birefringence.
In order to investigate whether the effective model can be used to extract the correct fibre inclinations, the retardation computed from Eq.\ (\ref{eq:phaseshift}) was compared to the retardation values derived from simulations using the macroscopic and the microscopic model (see Sec.\ \ref{sec:simulation_models}).
For that purpose, the retardation images were calculated for different fibre inclinations and different optical resolutions, respectively. 
For a better comparison between the retardation values of the single fibre and the fibre bundle, only the pixel in the centre of each (downsampled) retardation image was considered for evaluation. If pixels at other locations had been chosen, the retardation values of the single fibre would have been influenced by boundary effects that do not exist for the fibre bundle or real brain tissue which are completely filled with fibres.
The retardation values from the centre of each downsampled retardation image were plotted against the corresponding inclination angle, yielding a \textit{retardation curve} for each downsampling step.
The retardation curves were compared to the normalized retardation curve of the effective model (cf.\ Eq.\ (\ref{eq:phaseshift})), in the following referred to as \textit{theoretical curve}: $\vert\sin\delta\vert = \lvert \sin \left( (\pi/2) \cos^2\alpha \right) \rvert$.

To be able to compare different retardation curves, the retardation was normalized for each curve with the maximum retardation value, respectively:
\begin{align}
\left\vert\sin\hat{\delta}\right\vert = \sin \left( \frac{\pi}{2} \frac{\delta} {\delta_{\text{max}}} \right).
\label{eq:normalisation}
\end{align}

As only birefringent material (mainly myelin) is responsible for the phase shift in Eq.\ (\ref{eq:phaseshift}), $t$ describes not the thickness of the whole brain section but rather the \textit{local myelin thickness} $t_{\text{m}}$, i.\,e.\ the combined thickness of myelin sheaths along the optical path.
Due to the inhomogeneity of brain tissue, the local myelin density of a brain section is less than 100\,\%, i.\,e.\ the maximum possible retardation is $\vert\sin \left(\delta_{\alpha=0^{\circ},\text{max}}\right)\vert < 1$. If the inclination is calculated under the assumption that the maximum possible retardation equals $1$, the inclination angle will be overestimated. In order to obtain a more precise estimation of the inclination angle, a so-called \textit{myelin density correction} was applied to the downsampled retardation images:

In the case of the macroscopic model, in which the optic axes within one fibre have the same orientations, $\delta$ scales linearly with $t_{\text{m}}$. In the case of the microscopic model, in which the optic axes within one fibre have different inclination angles, the upper limit of $\delta$ scales linearly with $t_{\text{m}}$ as long as the optic axes of neighbouring myelin voxels have similar orientations (see Appx.\ \ref{appx:PhaseShift_MyelinDensity}).
Thus, the dependence on the myelin density can be eliminated to the greatest possible extent by multiplying the phase shift $\delta$ with a correction factor $(t/t_{\text{m}})$:
\begin{align}
\vert\sin\left(\delta_{\text{corr}}\right)\vert = \left| \sin \left( \frac{t}{t_{\text{m}}} \, \delta \right) \right| .
\label{eq:myelin_density_correction}
\end{align}
In order to apply the myelin density correction to the downsampled retardation images, $t_{\text{m}}$ was replaced by the combined thickness of myelin voxels along the optical  path (after applying the Gaussian filter and resampling). The resulting retardation images were normalized according to Eq.\ (\ref{eq:normalisation}), yielding $\vert \sin (\hat{\delta}_{\text{corr}}) \vert$.


\section{Simulation results}
\enlargethispage{0.5cm}

\subsection{Comparison of analytical and numerical solution}
\label{sec:results_comparison}

\begin{figure}[htbp]
\centering
\includegraphics[width=0.8 \textwidth]{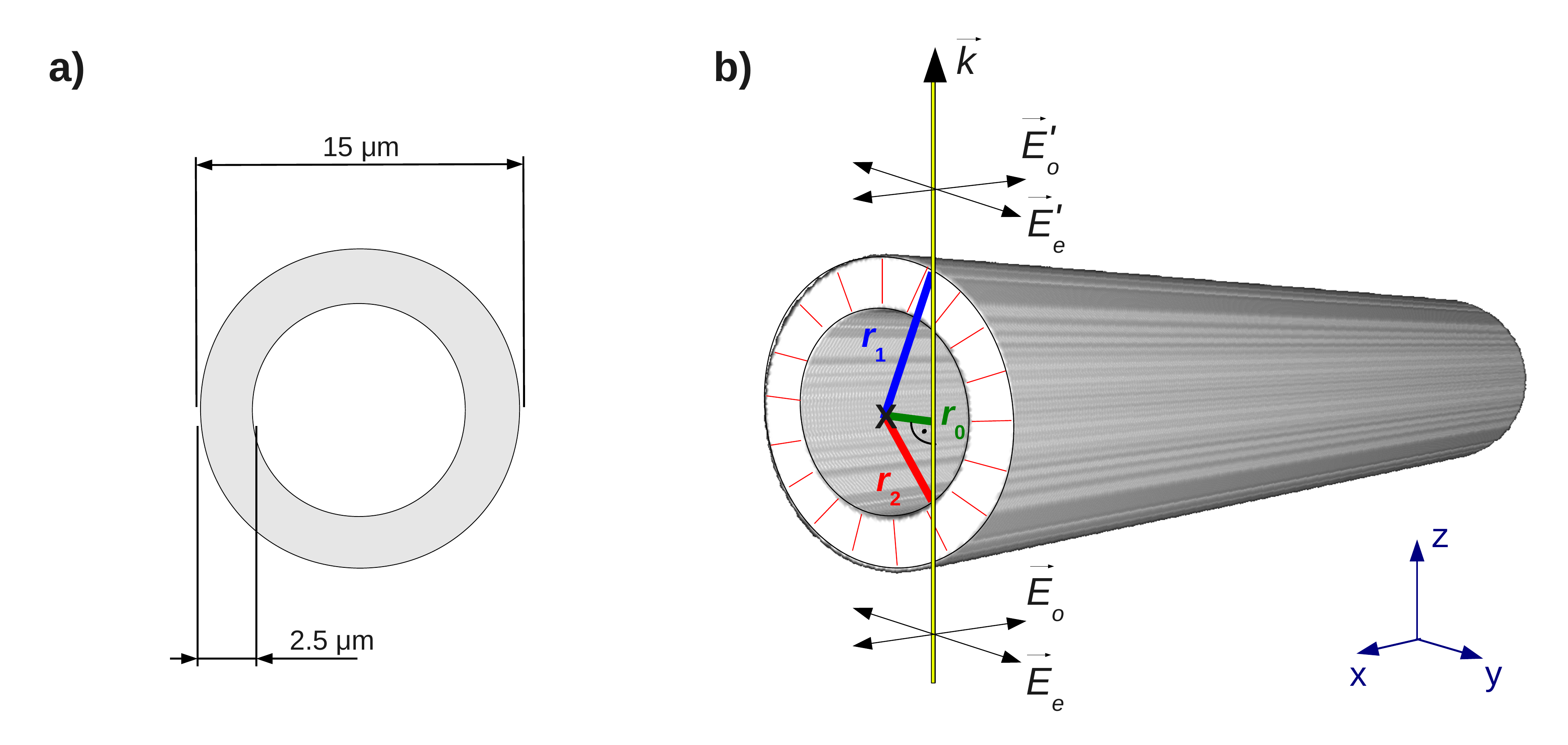}
\caption{\textbf{(a)} Dimensions of the simulated fibre (cross-sectional view). \textbf{(b)} Simulation model for comparison with the analytical solution: A horizontal fibre is simulated with outer radius $r_1 = 7.5$\,\textmu m, inner radius $r_2 = 5$\,\textmu m, and radial optic axes. The light is incident perpendicular to the fibre axis at distance $r_0$. The electric field vector of the ordinary wave ($\vec{E_{o}}$) is oriented parallel to the longitudinal axis of the fibre. The electric field vector of the extraordinary wave ($\vec{E_{e}}$) is oriented perpendicular to the fibre axis.}
\label{fig:simulated_fibre}
\end{figure}

To estimate the accuracy of the simulation results for the microscopic model, a single fibre with radial optic axes and perpendicularly incident light (see Fig.\ \ref{fig:simulated_fibre}b) was generated for different voxel sizes ($\Delta x_{\text{sim}}$) and the numerically computed phase difference between extraordinary and ordinary wave ($\Delta \varPhi_{\text{num}}$) was compared to the analytical solution ($\Delta \varPhi_{\text{ana}}$). 

Assuming that reflection and refraction effects can be neglected so that associated extraordinary and ordinary wave follow the same pathway, \textsc{Bear} and \textsc{Schmidt} derived an analytical expression for the phase difference \cite{bear36}:
\begin{align}
\Delta \varPhi_{\text{ana}} = \frac{2\pi}{\lambda}\varGamma \approx \frac{4\pi}{\lambda} \, r_0 \, \Delta n \left( \arccos \left( \frac{r_0}{r_1} \right) - \arccos \left( \frac{r_0}{r_2} \right) \right),
\label{eq:phasediff_ana}
\end{align}
where $\varGamma$ is the optical path length difference between extraordinary and ordinary wave, $r_1$ the radius of the whole nerve fibre (outer cylinder), $r_2$ the radius of the non-birefringent axon (inner cylinder) and $r_0$ the distance at which the light is incident perpendicular to the fibre axis (see Fig.\ \ref{fig:simulated_fibre}b). 

In order to compute $\Delta \varPhi_{\text{num}}$, the propagation of ordinary and extraordinary wave were simulated separately: In the case of the ordinary wave, the light is polarized parallel to the longitudinal axis of the fibre. In the case of the extraordinary wave, the light is polarized perpendicular to the longitudinal axis of the fibre (see Fig.\ \ref{fig:simulated_fibre}b). The phase for both the ordinary wave ($\varPhi_o$) and the extraordinary wave ($\varPhi_e$) was calculated from the corresponding electric field vector $\vec{E_T}$ in Eq.\ (\ref{eq:E_sim}):
\begin{align}
\varPhi = \arctan \left( \frac{\operatorname{Im}(\lvert\vec{E_T}\rvert)}{\operatorname{Re}(\vert\vec{E_T}\rvert)} \right).
\end{align}
The numerically computed phase difference $\Delta \varPhi_{\text{num}} = \varPhi_e - \varPhi_o$ was evaluated at various distances $0 < r_0 < 5$\,\textmu m away from the centre of the fibre and compared to the analytical solution given in Eq.\ (\ref{eq:phasediff_ana}), with $r_1 = 7.5$\,\textmu m, $r_2 = 5$\,\textmu m, $\lambda = 525$\,nm and $\Delta n = 0.001875$ (cf.\ Sec.\ \ref{sec:simulation_parameters}).
In order to study the impact of the spatial discretisation on the accuracy of the numerical solution, the simulation was performed for various voxel sizes $1.50$\,\textmu m $ > \Delta x_{\text{sim}} > 0.06$\,\textmu m. As a measure of consistency between the numerical and the analytical solution, the relative phase difference was calculated: $(\Delta \varPhi_{\text{ana}} - \Delta \varPhi_{\text{num}})/ \Delta \varPhi_{\text{ana}}$.

\begin{figure}[ht]
\centering
\includegraphics[width=1 \textwidth]{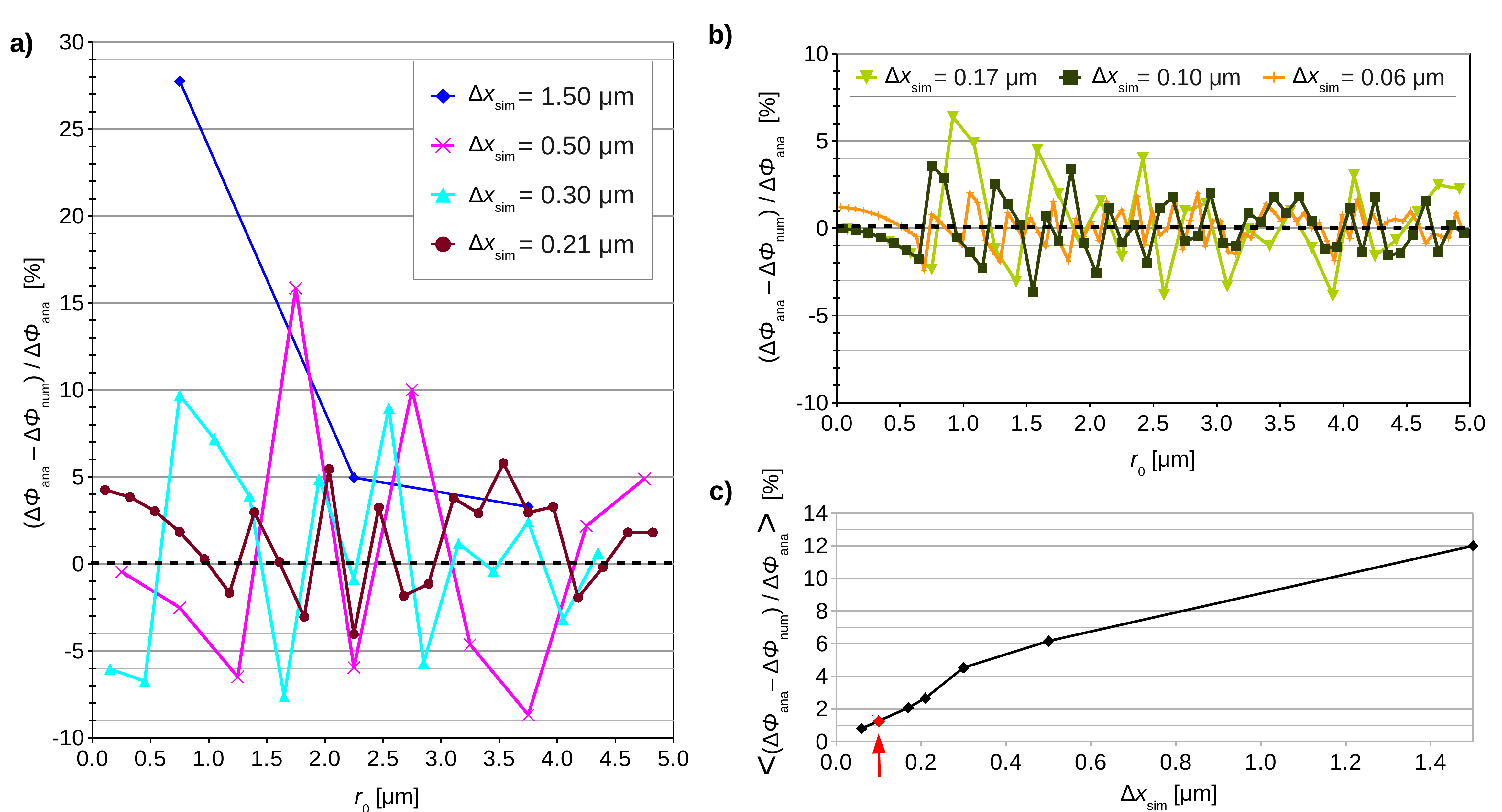}
\caption{\textbf{(a,b)} Relative difference between analytically and numerically calculated phase difference ($\Delta\varPhi_{\text{ana}}$ and $\Delta\varPhi_{\text{num}}$) for various voxel sizes $\Delta x_{\text{sim}}$, evaluated at different distances $r_0$ away from the centre of the fibre. For reasons of clarity, the results are presented in two diagrams: (a) $\Delta x_{\text{sim}} = 1.50$--$0.21$\,\textmu m, (b) $\Delta x_{\text{sim}} = 0.17$--$0.06$\,\textmu m. The dashed black lines indicate the point at which the numerical values match the analytical solution.
\textbf{(c)} Mean absolute relative phase difference plotted against the voxel size $\Delta x_{\text{sim}}$. The arrow indicates the voxel size ($\Delta x_{\text{sim}} = 0.1$\,\textmu m) that is chosen for the fibre simulations.}
\label{fig:phasediff_sim}
\end{figure}

Figures \ref{fig:phasediff_sim}a and \ref{fig:phasediff_sim}b show the relative phase difference plotted against $r_0$ for various voxel sizes $\Delta x_{\text{sim}}$.
As can be seen, the numerical solution fluctuates around the analytical solution for voxel sizes of 0.5\,\textmu m and less. With smaller voxel sizes, the numerical solution approaches the analytical solution (indicated by the dashed black line). This behaviour is especially evident when considering the mean of the absolute relative phase difference for each voxel size (see Fig.\ \ref{fig:phasediff_sim}c):
For a voxel size of $\Delta x_{\text{sim}} = 1.5$\,\textmu m (corresponding to one tenth of the fibre diameter), the mean absolute relative phase difference is about 12\,\%. For $\Delta x_{\text{sim}} = 0.5$\,\textmu m, it is about 6\,\% and for $\Delta x_{\text{sim}} = 0.06$\,\textmu m, it is only 0.8\,\%.
This demonstrates that the simulation tool produces correct results. 

As a good compromise between computation time and accuracy, all following fibre simulations were performed with a voxel size of $\Delta x_{\text{sim}} = 0.1$\,\textmu m (corresponding to 1/150 of the fibre diameter). For this voxel size, the relative phase difference is no more than 4\,\% (see Fig.\ \ref{fig:phasediff_sim}b) and the mean relative phase difference is about 1.3\,\% (see Fig.\ \ref{fig:phasediff_sim}c).


\subsection{Simulation of a single fibre}
\label{sec:simulation_single_fibre}

In a preliminary study, the limitations of the effective model of uniaxial negative birefringence were first studied for a straight single fibre. The fibre was simulated according to both the macroscopic and the microscopic model with different inclination angles (\mbox{$\alpha = 0^{\circ},\,10^{\circ},\, \dots,\,90^{\circ}$}) and different optical resolutions. The dimensions of the fibre and the other simulation parameters were chosen as described in Sec.\ \ref{sec:simulation_parameters}. 
The retardation curves were calculated from the downsampled retardation images (without/with myelin density correction) and normalized as described in Sec.\ \ref{sec:simulation_RetardationCurve}. An example of downsampled and corrected retardation images can be found in Appx.\ \ref{appx:RetImages}. 

\begin{figure}[htbp]
\centering
\includegraphics[width=1 \textwidth]{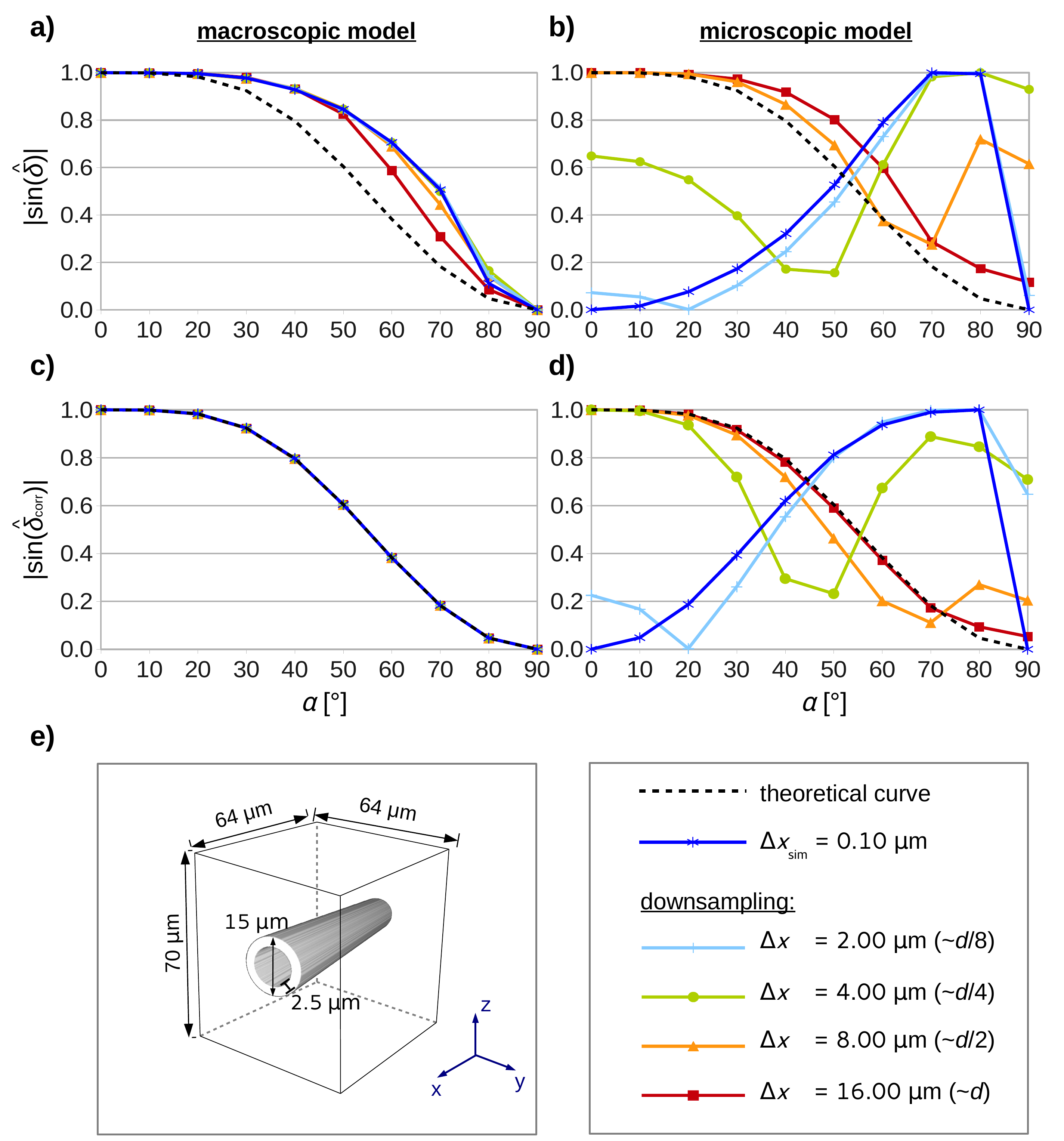}
\caption{\textbf{(a-d)} Normalized retardation curves of a straight single fibre simulated according to the macroscopic model (a,c) and the microscopic model (b,d) for different optical resolutions. Graphs (a,b) show the uncorrected retardation curves, graphs (c,d) show the retardation curves after the myelin density correction. For reasons of clarity, only selected graphs are shown. The legend indicates the pixel sizes of the retardation images from which the retardation curves have been calculated. The pixel size $\Delta x$ of the downsampled retardation images determines the parameters used for simulating the optical resolution (see Tab.\ \ref{tab:downsampling}). For better comparison, $\Delta x$ is also given in terms of the fibre diameter ($d=15$\,\textmu m). \textbf{(e)} Dimensions of the simulated single fibre.}
\label{fig:singlefibre_retcurve}
\end{figure}

Figure \ref{fig:singlefibre_retcurve} shows the dimensions of the simulated single fibre and the corresponding retardation curves (continuous lines) for both simulation models and different optical resolutions (according to Tab.\ \ref{tab:downsampling}). The theoretical retardation curve of the effective model is indicated by a dashed black line.
In the case of the macroscopic model, the uncorrected retardation curves  (see Fig.\ \ref{fig:singlefibre_retcurve}a) are already very similar to the theoretical curve for all investigated optical resolutions. After the myelin density correction (see Fig.\ \ref{fig:singlefibre_retcurve}c), all retardation curves match the theoretical curve exactly, independently of the optical resolution.
In the case of the microscopic model (see Figs.\ \ref{fig:singlefibre_retcurve}b and \ref{fig:singlefibre_retcurve}d), the retardation curves for a pixel size much smaller than the fibre diameter ($\Delta x < 2$\,\textmu m) are inverted as compared to the theoretical curve for $\alpha < 90^{\circ}$, i.\,e.\ the microscopic and the macroscopic model yield totally different results. For intermediate pixel sizes ($2$\,\textmu m $\leq \Delta x \leq 8$\,\textmu m), the retardation curves are non-monotonic, i.\,e.\ the assignment of the inclination angle is ambiguous. Finally, for pixel sizes larger than the fibre diameter ($\Delta x = 16$\,\textmu m), the uncorrected retardation curve (see Fig.\ \ref{fig:singlefibre_retcurve}b) is similar to the theoretical curve. 
After the myelin density correction  (see Fig.\ \ref{fig:singlefibre_retcurve}d), the retardation curve matches the theoretical curve almost exactly.


\subsection{Simulation of a fibre bundle}
\label{sec:simulation_fibre_bundle}

\begin{figure}[htbp]
\centering
\includegraphics[width=1 \textwidth]{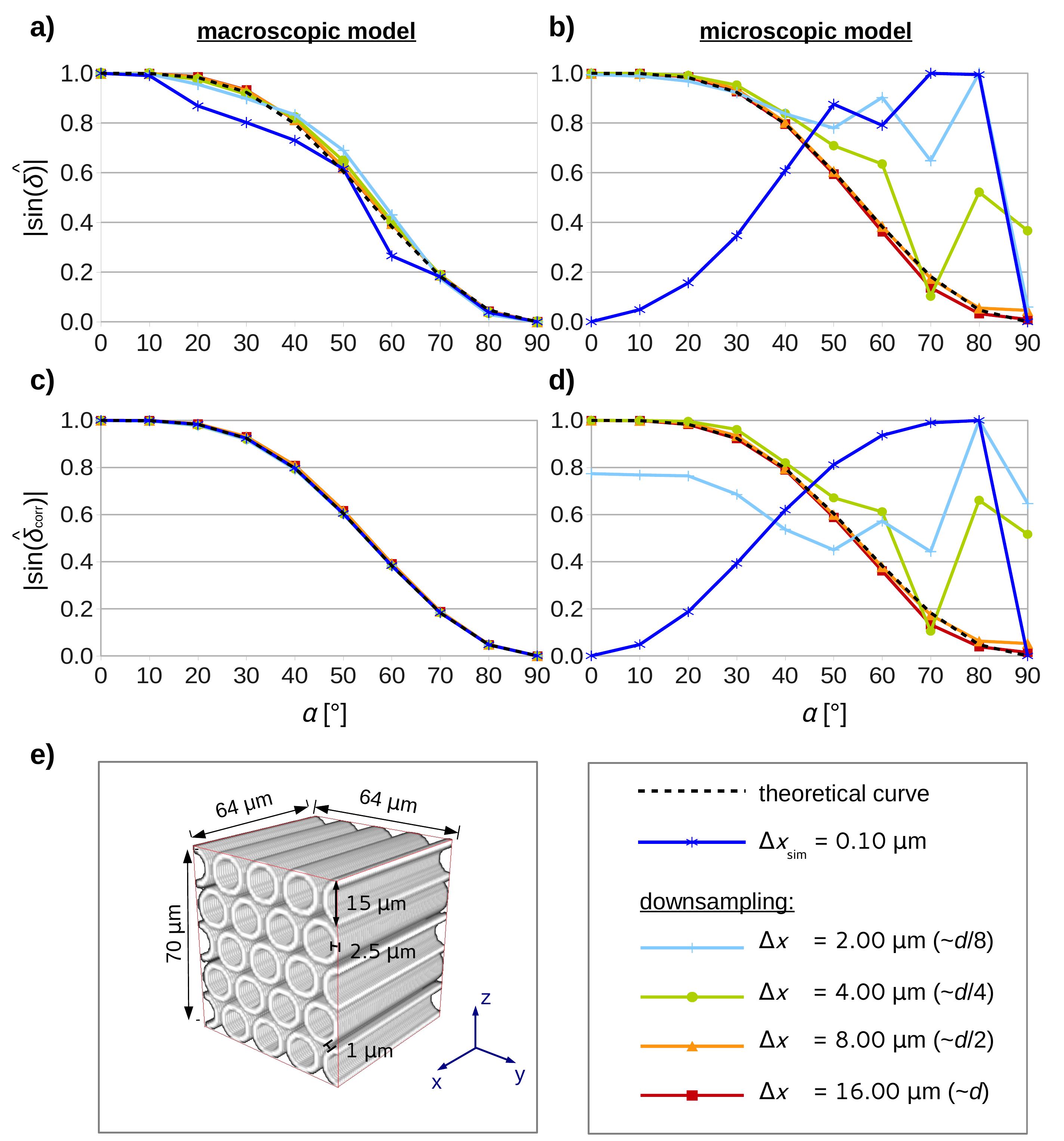}
\caption{\textbf{(a-d)} Normalized retardation curves of a hexagonal fibre bundle simulated according to the macroscopic model (a,c) and the microscopic model (b,d) for different optical resolutions. Graphs (a,b) show the uncorrected retardation curves, graphs (c,d) show the retardation curves after the myelin density correction. For reasons of clarity, only selected graphs are shown. The legend indicates the pixel sizes of the retardation images from which the retardation curves have been calculated. The pixel size $\Delta x$ of the downsampled retardation images determines the parameters used for simulating the optical resolution (see Tab.\ \ref{tab:downsampling}). For better comparison, $\Delta x$ is also given in terms of the fibre diameter ($d=15$\,\textmu m). \textbf{(e)} Dimensions of the simulated fibre bundle.}
\label{fig:fibrebundle_retcurve}
\end{figure}

In brain tissue, nerve fibres are usually organised in hexagonal close-packed fibre bundles \cite{quarles06}. In order to investigate the effect of fibre bundles on the 3D-PLI signal, a hexagonal bundle of straight parallel fibres with an inter-fibre spacing of 1\,\textmu m was simulated (see Fig.\ \ref{fig:fibrebundle_retcurve}e). In order to obtain comparable results, the same dimensions and simulation parameters were chosen as for the single fibre.

Figure \ref{fig:fibrebundle_retcurve} shows the normalized retardation curves for both simulation models and different optical resolutions (according to Tab.\ \ref{tab:downsampling}).
The (downsampled) retardation images that were used to compute the corrected retardation curves of the microscopic model are shown in Appx.\ \ref{appx:RetImages}.

In the case of the macroscopic model, the uncorrected retardation curves (see Fig.\ \ref{fig:fibrebundle_retcurve}a) are very similar to the theoretical retardation curve of the effective model (dashed black line) for all investigated optical resolutions. As compared to the retardation curves of the single fibre (see Fig.\ \ref{fig:singlefibre_retcurve}a), the retardation curves of the fibre bundle are closer to the theoretical curve. After the myelin density correction (see Fig.\ \ref{fig:fibrebundle_retcurve}c), the curves are almost identical. 
In the case of the microscopic model, the uncorrected retardation curves (see Fig.\ \ref{fig:fibrebundle_retcurve}b) are also closer to the theoretical curve as compared to the uncorrected retardation curves of the single fibre (see Fig.\ \ref{fig:singlefibre_retcurve}b). The myelin density correction (see Fig.\ \ref{fig:fibrebundle_retcurve}d) makes only a small difference, especially for low optical resolutions.
For the simulated fibre bundle, the transition between the microscopic and the macroscopic model already occurs for pixel sizes larger than the fibre radius ($\Delta x \geq 8$\,\textmu m).


\section{Discussion}

In 3D-PLI, the fibre orientations are derived under the assumption that the brain tissue can (locally) be described as a homogeneous and uniaxial birefringent material with the optic axis indicating the predominant fibre direction. Furthermore, the density of myelinated fibres is assumed to be the same for the whole brain section. In this paper, the limitations of this effective birefringence model have been studied for the first time. For that purpose, a single fibre and a hexagonal fibre bundle (with diameters $d$) were simulated based on the Jones matrix calculus, employing a microscopic and a macroscopic model of birefringence and different optical resolutions (defined by the pixel size $\Delta x$ as given in Tab.\ \ref{tab:downsampling}).

The transition between the two models is apparent when analysing the retardation curves: For high optical resolutions ($\Delta x << d$), the radial optic axes of the microscopic model are resolved. In this case, the optic axes are oriented perpendicular to the longitudinal fibre axis so that the retardation curves are inverted as compared to the macroscopic model and fibres with high inclination angles are interpreted as flat fibres. The zero retardation value for $\alpha=90^{\circ}$ is an artifact arising from the fact that the retardation is evaluated at the centre of the retardation image which -- in the case of vertical fibres -- contains no myelin (cf.\ Fig.\ \ref{fig:fibrebundle_retimage}, upper right corner).
For intermediate optical resolutions ($\Delta x < d$), there is a transition zone between the microscopic and the macroscopic model so that an unambiguous assignment between retardation and inclination is not possible. For sufficiently low optical resolutions (single fibre: $\Delta x > d$; fibre bundle: $\Delta x > d/2$), the microscopic and the macroscopic model yield similar results (see Figs.\ \ref{fig:singlefibre_retcurve}d and \ref{fig:fibrebundle_retcurve}d) so that the effective model of uniaxial negative birefringence can be used to compute the fibre inclinations. 

Thus, for the simulated fibre bundle (consisting of five fibre layers with $d=15$\,\textmu m), the effective model can be used to interpret LAP measurements ($\Delta x_{\text{LAP}} = 64$\,\textmu m $> d/2$), but not to interpret PM measurements ($\Delta x_{\text{PM}} = 1.33$\,\textmu m $< d/2$). However, the diameters of the simulated fibres represent an upper estimate of typical fibre diameters in the human brain. The diameters of myelinated nerve fibres range from $0.3$ to $15$\,\textmu m \cite{aboitiz92,longstaff00,morell,hildebrand92} and the majority of the fibres (e.\,g.\ 80\,\% in the corpus callosum \cite{aboitiz92}) have diameters of 1\,\textmu m or below so that the condition $\Delta x_{\text{PM}} > d/2$ is still fulfilled.
In addition, fibre diameters much smaller than $15$\,\textmu m implicate that the measured brain section (with thickness 70\,\textmu m) contains much more fibre layers than the simulated fibre bundle. A comparison between the simulated single fibre and the fibre bundle suggests that the more fibre layers are located along the optical path, the smaller is the minimum pixel size for which the effective model is still valid.
To verify this hypothesis, the limitations of the effective model should also be studied in terms of the number of fibre layers along the optical path.
However, a larger number of fibre layers also increases the probability that fibres with different spatial orientations are measured within the same volume, which poses a major challenge for 3D-PLI \cite{dohmen15}. In future studies, the limitations of the effective model should therefore also be investigated for non-parallel fibre structures.

The simulations have shown that -- in regions with parallel fibre structures -- the effective model of uniaxial negative birefringence is valid for the employed optical set-ups. For imaging systems with very high optical resolutions, the effective model needs to be reconsidered.
Even if the optical resolution is too high to extract the correct fibre inclinations, 3D-PLI remains a valuable neuroimaging technique as the image contrasts of transmittance and retardation still provide detailed structural information on the two-dimensional nerve fibre architecture in large histological brain sections. 

The effective model that is currently used for the data analysis in 3D-PLI does not only assume parallel fibre structures, but also a uniform myelin density. 
The simulations have shown that the retardation signal is considerably influenced by the myelin density, which impairs the reconstructed fibre orientations. 
It could be demonstrated that the estimation of the fibre inclination is considerably improved by the myelin density correction which incorporates the local myelin thickness of the examined tissue into the calculation of the inclination angle. While the correction has a large effect on the retardation curves of the single fibre, the effect is smaller for the fibre bundle which is much more homogeneous than the single fibre. Thus, the myelin density correction is especially useful for regions with an inhomogeneous density of myelinated nerve fibres (e.\,g.\ for transition zones between white and grey matter).
In the case of the microscopic model, the correction does not work as well as for the macroscopic model because the retardation also depends on the direction of the radially oriented optic axes in the myelin sheath, but it is still a considerable improvement.
In order to incorporate the myelin density correction into the 3D-PLI signal analysis, the local myelin thickness $t_{\text{m}}$ of the sample needs to be determined.
The intensity values of the transmittance image seem to be a good measure of the local myelin thickness in brain tissue \cite{axer11_2}.

The purpose of this study was to explore and understand the most dominant effects that generate the birefringence signals in 3D-PLI. 
To fully understand the physical processes behind 3D-PLI and to improve the interpretation of the reconstructed fibre orientations, a direct comparison between simulation and experiment is required.
The long-term aim should be to develop a simulation tool of 3D-PLI that considers all relevant effects needed for reproducing the experimental results. To this end, the simulation model should be extended step by step and the relevant effects should be identified.

Although the simulations show that the simplified microscopic model can already be used to explain the effective negative birefringence of parallel nerve fibres, future studies should include the positive birefringence of the axon and investigate how this modification changes the transition between the microscopic and the macroscopic model.

So far, only straight and parallel fibres have been investigated.
To provide more realistic fibre models, the fibres should be simulated with varying fibre diameters, myelin sheath thickness and spatial orientations. As fibres with different spatial orientations pose a major challenge for 3D-PLI \cite{dohmen15}, future studies should focus on investigating inhomogeneous, non-parallel fibre structures. To enable a direct comparison with the experiment, the simulated fibre configurations should be based on experimentally determined fibre structures.

In addition to a more realistic fibre model, the propagation of light should also be simulated more realistically.
In this study, the incident light was described by a parallel beam of light. However, in the experiment, the employed light source emits diffusive light, i.\,e.\ the sample is illuminated by light with slightly different angles of incidence. As the measured birefringence signals depend on the angle between the light wave and the nerve fibres, a non-zero angle of incidence changes the retardation curves. 
For the LAP, which has a small numerical aperture, the effect can presumably be neglected. However, for systems with higher optical resolutions and higher numerical apertures, the effect of a Gaussian distribution of incident angles should be investigated further.

Moreover, the simulations were based on the Jones matrix calculus which is only applicable to completely polarized and coherent light. As the light source emits incoherent light and the polarizers are not perfect, the Jones matrices should be replaced by M\"{u}ller matrices \cite{mueller43} which enable to study partially polarized and incoherent light.

Finally, the assumption of a linear optical pathway is a great simplification. The refractive index of the myelin sheath is higher than the refractive indices of the inner axon and the surrounding tissue \cite{vidal80,duck90,beuthan96} which will cause refraction/reflection at the interfaces and scattering of light. In future studies, the effects of refraction and scattering on the measured birefringence signal should be investigated in more detail. As the used simulation tool (SimPLI) is based on a matrix calculus, other simulation approaches will be required to investigate such non-linear pathways.


\section{Conclusion}

In this study, we laid a theoretical foundation for 3D-PLI. The effective model of uniaxial negative birefringence, which is currently used to compute the nerve fibre orientations from experimental data, has been validated for the first time. Using simulations based on the Jones matrix calculus, we have shown that the effective model can be used for the employed optical set-ups, i.\,e.\ as long as the polarimeter does not resolve structures smaller than the diameter of single nerve fibres.
The developed Jones matrix formalism for simulating 3D-PLI has proven to be a powerful tool to gain a deeper theoretical understanding of the physical processes behind 3D-PLI and to better interpret the experimental data. The simulations enable not only to validate the computational model of the fibre reconstruction, but also to optimise the experimental set-up and the measurement method.


\section*{Competing interests}
The authors declare that they have no competing interests.


\section*{Authors' contributions}
M. Menzel substantially contributed to the conception and design of the study as well as to the acquisition, analysis and interpretation of the simulated data. She carried out the simulations as well as the analytical calculations and drafted the manuscript.
K. Michielsen participated in the design of the study, contributed to theoretical considerations and to the interpretation of the simulated data, and helped draft the manuscript.
H. De Raedt contributed to the interpretation of the simulated data, to the theoretical
considerations and to the revision of the manuscript.
J. Reckfort conducted experimental measurements, helped transfer the measurement results to
the simulation and revised the manuscript.
K. Amunts contributed to the anatomical content of the study and to the revision of the
manuscript.
M. Axer coordinated the study, participated in the conception and design, contributed to the
analysis and interpretation of the simulated data, and helped draft the manuscript.
All authors read the final manuscript and gave final approval for publication.


\section*{Aknowledgements}
The authors thank Melanie Dohmen for the introduction to the simulation software SimPLI.


\section*{Funding statement}
This work was supported by the Helmholtz Association portfolio theme ''Supercomputing and Modeling for the Human Brain`` and by the European Union Seventh Framework Programme (FP7/2007--2013) under grant agreement no.\ 604102 (Human Brain Project).


\bibliographystyle{unsrt}
\bibliography{PAPER}


\appendix

\section{Derivation of the phase shift}
\label{appx:derivation_PhaseShift}

When polarized light passes through the birefringent brain section, it is split into an ordinary and an extraordinary wave which both experience different refractive indices. The refractive index $n_e$ that the extraordinary wave experiences when passing through the birefringent tissue under an angle $\theta$ with respect to the optic axis, is given by \cite{born}:
\begin{align}
\frac{1}{n_e(\theta)^2} &= \frac{1}{n_o^2} \cos^2\theta + \frac{1}{n_E^2} \sin^2\theta 
\label{eq:1/n_e(theta)2} \\
\Leftrightarrow \frac{1}{n_o^2} - \frac{1}{n_e(\theta)^2} 
&= \left( \frac{1}{n_o^2} - \frac{1}{n_E^2} \right) \sin ^2 \theta ,
\label{eq:1/no2-1/ne2}
\end{align}
where $n_o$ is the ordinary refractive index and $n_E \equiv n_e(\theta = 90^{\circ})$ the principal extraordinary refractive index of the brain tissue.

The birefringence of biological tissue ($\Delta n  = n_E - n_o = 10^{-3} ... 10^{-2}$ \cite{ghosh11}) is small as compared to the values of the refractive indices $n_o$ and $n_E$ ($n=1.3$--$1.5$ \cite{beuthan96}). Therefore, a Taylor expansion can be applied to the function
\begin{align}
f(\Delta n) \equiv \frac{1}{n_o^2} - \frac{1}{n_E^2} = \frac{1}{n_o^2} - \frac{1}{(n_o + \Delta n)^2}
\end{align}
in $\Delta n = 0$: 
\begin{align}
f(\Delta n) = \sum\limits_{l=0}^{\infty} \frac{f^{(l)}(0)}{l!} (\Delta n)^l = f(0) + f'(0)\,\Delta n + ... = 0 + \frac{2}{n_o^3} \Delta n + ...
\label{eq:TaylorExpansion}
\end{align}
The same expansion can be done for $\left(1/n_o^2 - 1/n_e(\theta)^2\right)$ in $(\Delta n(\theta) = n_e(\theta) - n_o \ll 1)$. With these Taylor expansions, Eq.\ (\ref{eq:1/no2-1/ne2}) can be written as:
\begin{align}
\Delta n(\theta) \approx \Delta n \, \sin ^2\theta.
\label{eq:Delta_n}
\end{align}

Choosing a coordinate system in which the light propagates in the z-direction and the brain tissue lies in the xy-plane, the optic axis (oriented in the direction of the nerve fibres) makes an angle $\theta$ with the z-axis, i.\,e.\ the out-of-plane inclination angle of the fibre is $\alpha = 90^{\circ} - \theta$. With this definition follows: $\Delta n (\theta) \approx \Delta n \,\cos^2\alpha$.

Thus, when the light passes through a brain section of thickness $t$, the extraordinary wave experiences a phase shift with respect to the ordinary wave which depends on the inclination angle of the optic axis:
\begin{align}
\delta = \frac{2\pi}{\lambda} t \, \Delta n(\theta) \approx \frac{2\pi}{\lambda} t \, \Delta n \cos^2\alpha.
\end{align}
This is the formula of the phase shift as given in Eq.\ (\ref{eq:phaseshift}).


\section{Derivation of the downsampling parameters}
\label{appx:OpticalResolution}

\begin{figure}[htbp]
\centering
\includegraphics[width=1 \textwidth]{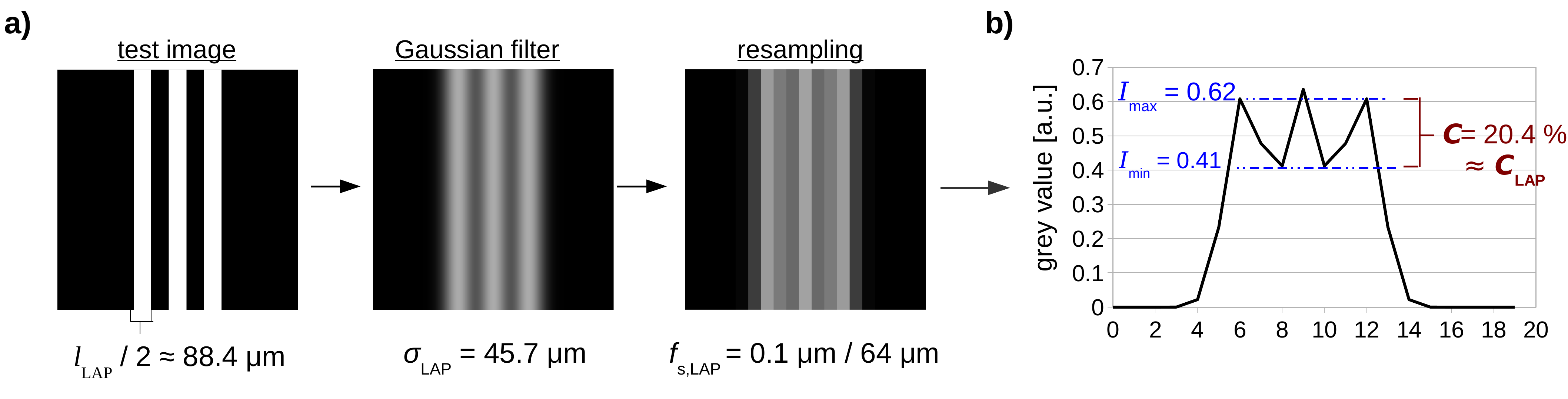}
\caption{
\textbf{(a)} Downsampling of a test image (grey values: black $=0$, white $=1$): A Gaussian filter with standard deviation $\sigma_{\text{LAP}}$ and resampling with sampling factor $f_{\text{s,LAP}}$ are applied to the test image. \textbf{(b)} Line profile of the downsampled test image: The determined contrast $\mathcal{C} = (I_{\text{max}} - I_{\text{min}})/(I_{\text{max}} + I_{\text{min}})$ matches approximately the contrast $\mathcal{C}_{\text{LAP}}$ obtained from experimental measurements.}
\label{fig:downsampling}
\end{figure}

In previous measurements, the optical resolution of the LAP was investigated by employing a \textit{USAF} test chart which contains line pairs (lp) with different spacings \cite{reckfort2013}. From the measured line intensity profiles, the Michelson contrast $\mathcal{C}$ was computed:
\begin{align}
\mathcal{C} = \frac{I_{\text{max}} - I_{\text{min}}}{I_{\text{max}} + I_{\text{min}}},
\end{align}
where $I_{\text{max}}$ corresponds to the mean intensity of the maxima and $I_{\text{min}}$ to the mean intensity of the (local) minima in the line intensity profile (cf.\ Fig.\ \ref{fig:downsampling}b).
The largest number of line pairs per millimetre that can just be resolved (according to the Rayleigh criterion) was determined to be 5.66\,lp/mm, which corresponds to a width per line pair of $l_{\text{LAP}} = 176.7$\,\textmu m and a contrast of $\mathcal{C}_{\text{LAP}} = 20.1\,\%$. A width per line pair of $157.5$\,\textmu m yields a Michelson contrast of $11.2\,\%$.

According to these measurement results, a test image with three lines (pixel size: 0.1\,\textmu m) and a line width of $l_{\text{LAP}}/2 \approx 88.4$\,\textmu m was created, and the downsampling procedure (Gaussian filter and resampling) was applied to the test image (see Fig.\ \ref{fig:downsampling}a). The sampling factor was calculated by dividing the pixel size of the test image by the pixel size of the LAP: $f_{\text{s,LAP}}=\text{0.1\,\textmu m} / \text{64\,\textmu m}$.  To reproduce the measured contrast of the line intensity profile (see Fig.\ \ref{fig:downsampling}b), a Gaussian filter with a standard deviation of $\sigma_{\text{PM}} = 45.7$\,\textmu m was applied. To avoid boundary effects and ensure a symmetric line intensity profile, the dimensions of the image ($1216$\,\textmu m $\times 1216$\,\textmu m) were chosen such that the downsampled image consists of an odd number of pixels ($19$\,px $\times 19$\,px).

Based on the determined parameters for the LAP ($\Delta x_{\text{LAP}}$, $\sigma_{\text{LAP}}$, $f_{\text{s,LAP}}$), downsampling parameters for imaging systems with other optical resolutions (see Tab.\ \ref{tab:downsampling}) were derived: The ratio between the pixel size of the LAP image and the determined standard deviation of the two-dimensional Gaussian filter is $\sigma_{\text{LAP}}/\Delta x_{\text{LAP}} = 45.7$\,\textmu m $/64$\,\textmu m $\approx 0.714$. Analogous measurements of the PM yield a similar ratio between pixel size and standard deviation \cite{reckfort2013}. Assuming that this ratio is the same for all simulated imaging systems, the standard deviation of the two-dimensional Gaussian filter was calculated from the pixel size $\Delta x$ of the resulting downsampled image:
\begin{align}
\sigma = 0.714 \, \Delta x.
\end{align}
After applying the Gaussian filter, the synthetic image series (with pixel size $\Delta x_{\text{sim}}$) was resampled with a sampling factor of
\begin{align}
f_{\text{s}} = \frac{\Delta x_{\text{sim}}}{\Delta x} = \frac{0.1 \text{\,\textmu m}}{\Delta x},
\end{align}
yielding a downsampled image series with pixel size $\Delta x$.


\section{Dependence of the phase shift on the local myelin thickness}
\label{appx:PhaseShift_MyelinDensity}

Each myelin voxel of a simulated nerve fibre is represented by the Jones matrix of a rotated wave retarder as defined in Eq.\ (\ref{eq:M_retarder}). 
Depending on what kind of model is used (macroscopic or microscopic), the retarder axis is either oriented parallel or radially to the fibre axis (see Fig.\ \ref{fig:simulation_method}b).


In the macroscopic model, the optic axes of the myelin voxels are all oriented in the fibre direction ($\varphi$, $\alpha$) so that the voxels can be described by the same Jones matrix $M_{\delta}(\beta)$ with phase shift $\delta$ and $\beta \equiv \varphi - \rho$. When the light propagates through $N$ voxels of myelin, the multiplication of the $N$ corresponding Jones matrices yields (using Eq.\ (\ref{eq:M_retarder}) and $R(\beta) \, R(-\beta) = \mathbb{I}$):
\begin{align}
\big(M_{\delta}(\beta)\big)^N  
&=
R(\beta)\,
	\begin{pmatrix} e^{\operatorname{i} \delta/2} &  0 			\\ 
                	0 	 &   e^{-\operatorname{i} \delta/2}
    \end{pmatrix}
	\,R(-\beta)\,\cdots\,
	R(\beta)\,
	\begin{pmatrix} e^{\operatorname{i} \delta/2} &  0 			\\ 
                	0 	 &   e^{-\operatorname{i} \delta/2}
    \end{pmatrix}
	\,R(-\beta) \notag \\
&=
R(\beta)\,
	\begin{pmatrix} e^{\operatorname{i} \delta/2} &  0 			\\ 
                	0 	 &   e^{-\operatorname{i} \delta/2}
    \end{pmatrix}^N
    \,R(-\beta) \notag\\
&=
R(\beta)\,
	\begin{pmatrix} e^{\operatorname{i} N\,\delta/2} &  0 			\\ 
                	0 	 &   e^{-\operatorname{i} N\,\delta/2}
    \end{pmatrix}
    \,R(-\beta) \notag\\
&=
M_{(N\,\delta)}(\beta).
\end{align}
Thus, the $N$ myelin voxels with thickness $\Delta t$ (along the optical path) and phase shift $\delta$ can be replaced by one myelin voxel with side length $(N\,\Delta t \equiv t_{\text{m}})$ and phase shift:
\begin{align}
\delta ' \equiv N\,\delta \overset{(\ref{eq:phaseshift})}{=} \frac{2\pi}{\lambda}\,\Delta n\,(N\,\Delta t)\,\cos^2\alpha
= \frac{2\pi}{\lambda}\,\Delta n\,t_{\text{m}}\cos^2\alpha.
\end{align}
In other words, the phase shift $\delta$ (and for small $\delta$ also the retardation $\vert \sin\delta \vert$) scales linearly with the combined thickness of myelin voxels $(N\,\Delta t)$, i.\,e.\ with the local myelin thickness $t_{\text{m}}$. 


In the microscopic model, the optic axes of the myelin voxels along the optical path all have different orientations ($\varphi_j$, $\alpha_j$), see Fig.\ \ref{fig:simulation_method}c. If the optic axes of neighbouring myelin voxels have a similar direction ($\varphi_2 - \varphi_1 \ll 1$ and $\alpha_2 - \alpha_1 \ll 1$), the multiplication of the $N$ Jones matrices of the voxels can be simplified.
For $\varphi_2 - \varphi_1 \ll 1$, one can define $\beta_2 - \beta_1 \equiv \eta_{21} \ll 1$ and the multiplication of a pair of rotation matrices yields:
\begin{align}
R(-\beta_2) \cdot R(\beta_1)
&=
	\begin{pmatrix} \cos(\beta_2)  &  \sin(\beta_2)  \\ 
                  - \sin(\beta_2)  &  \cos(\beta_2)
    \end{pmatrix} \,
	\begin{pmatrix} \cos(\beta_1)  &  - \sin(\beta_1)  \\ 
                    \sin(\beta_1)  &    \cos(\beta_1)
    \end{pmatrix} \notag \\
&=
	\begin{pmatrix} \cos(\beta_2 - \beta_1)  &  \sin(\beta_2 - \beta_1)  				\\- \sin(\beta_2 - \beta_1)  &  \cos(\beta_2 - \beta_1)
    \end{pmatrix} 
\approx
	\begin{pmatrix}  1  		   &  \eta_{21}  \\ 
                  - \eta_{21}  &  1
    \end{pmatrix},
    \label{eq:R1_R2}
\end{align}
using a first-order approximation in $\eta_{21}$.\\
The multiplication of two Jones matrices yields (with $M_j \equiv M_{\delta_j}(\beta_j)$):
\begin{align}
M_2 \cdot M_1 
&\overset{(\ref{eq:M_retarder})}{=} 
R(\beta_2)\,
	\begin{pmatrix} e^{\operatorname{i} \delta_2/2} &  0 			\\ 
                	0 	 &   e^{-\operatorname{i} \delta_2/2}
    \end{pmatrix}
	\,R(-\beta_2) \cdot
	R(\beta_1)\,
	\begin{pmatrix} e^{\operatorname{i} \delta_1/2} &  0 			\\ 
                	0 	 &   e^{-\operatorname{i} \delta_1/2}
    \end{pmatrix}
	\,R(-\beta_1) \notag \\
&\overset{(\ref{eq:R1_R2})}{\approx}
R(\beta_2)\,
	\begin{pmatrix} e^{\operatorname{i}(\delta_1 +\delta_2)/2}
				&	\eta_{21}\,e^{\operatorname{i}(\delta_2 -\delta_1)/2}		\\ -\eta_{21}\,e^{-\operatorname{i}(\delta_2 -\delta_1)/2}
				&   e^{-\operatorname{i}(\delta_1 +\delta_2)/2}
    \end{pmatrix}
    R(-\beta_1).
\end{align}
The multiplication of four Jones matrices yields (ignoring terms in the order of $\eta_{ji}^2$):
\begin{align}
M_4 \cdot M_3 \cdot M_2 \cdot M_1 
\approx R(\beta_4)\,
	\begin{pmatrix} 
e^{\operatorname{i}(\delta_1 +\delta_2 +\delta_3 +\delta_4)/2}	
& \eta' \\
\eta'' & e^{-\operatorname{i}(\delta_1 +\delta_2 +\delta_3 +\delta_4)/2}	
	\end{pmatrix}
    R(-\beta_1), 
\end{align}
where the elements of the secondary diagonal are given by:
\begin{align}
\eta' &= \,\,\,\,\eta_{21}\,e^{\operatorname{i}(\delta_4 +\delta_3 +\delta_2 -\delta_1)/2}
\,\,\,\,+ \eta_{32}\,e^{\operatorname{i}(\delta_4 +\delta_3 -\delta_2 -\delta_1)/2} 
\,\,\,\,+\eta_{43}\,e^{\operatorname{i}(\delta_4 -\delta_3 -\delta_2 -\delta_1)/2}, \\
\eta'' &= - \eta_{21}\,e^{-\operatorname{i}(\delta_4 +\delta_3 +\delta_2 -\delta_1)/2} 
-\eta_{32}\,e^{-\operatorname{i}(\delta_4 +\delta_3 -\delta_2 -\delta_1)/2} 
-\eta_{43}\,e^{-\operatorname{i}(\delta_4 -\delta_3 -\delta_2 -\delta_1)/2}.
\end{align}
If the number of myelin voxels (i.\,e.\ the number of matrices $M_j$) is small, the elements of the secondary diagonal in the resulting matrix can be neglected for $\eta_{jk} \ll 1$. 
If the number of myelin voxels is large, the arguments of the exponential functions in the secondary diagonal will take all possible values and cancel each other for $\eta_{jk} \approx \eta_{lm}\,\,\forall\,\,j,k,l,m$.
In both cases, the multiplication of $N$ Jones matrices yields:
\begin{align}
M_N \cdot M_{N-1} \cdots M_1 
\approx R(\beta_N)\,
	\begin{pmatrix} e^{\operatorname{i}(\delta_1 + \cdots + \delta_N)/2} &  0 	\\ 0 & e^{-\operatorname{i}(\delta_1 + \cdots + \delta_N)/2} 
    \end{pmatrix} \,
R(-\beta_1).
\end{align}
Thus, the $N$ myelin voxels with thickness $\Delta t$ and phase shift $\delta_j$ can be replaced by one myelin voxel with thickness $(N \Delta t = t_{\text{m}})$ and phase shift:
\begin{align}
	\delta'	= \sum_{j=1}^N \delta_j 
		\overset{(\ref{eq:phaseshift})}{=} \frac{2\pi}{\lambda} \Delta n \,\Delta t \sum_{j=1}^N \cos^2\alpha_j
		\leq \frac{2\pi}{\lambda} \Delta n \,t_{\text{m}},
\end{align} 
given that the optic axes of neighbouring myelin voxels have similar directions.

The analytical considerations have shown that the phase shifts of individual voxels add together in both simulation models: In the macroscopic model, the phase shift scales linearly with the local myelin thickness $t_{\text{m}}$. In the microscopic model, this is only true for the upper limit of the phase shift.
The dependence on the local myelin thickness is taken into account in the myelin density correction (see Sec.\ \ref{sec:simulation_RetardationCurve}).


\newpage
\section{Retardation images of the fibre bundle}
\label{appx:RetImages}

\addtolength{\textheight}{-8cm}

\begin{figure}[htbp]
\centering
\includegraphics[width=0.83 \textwidth]{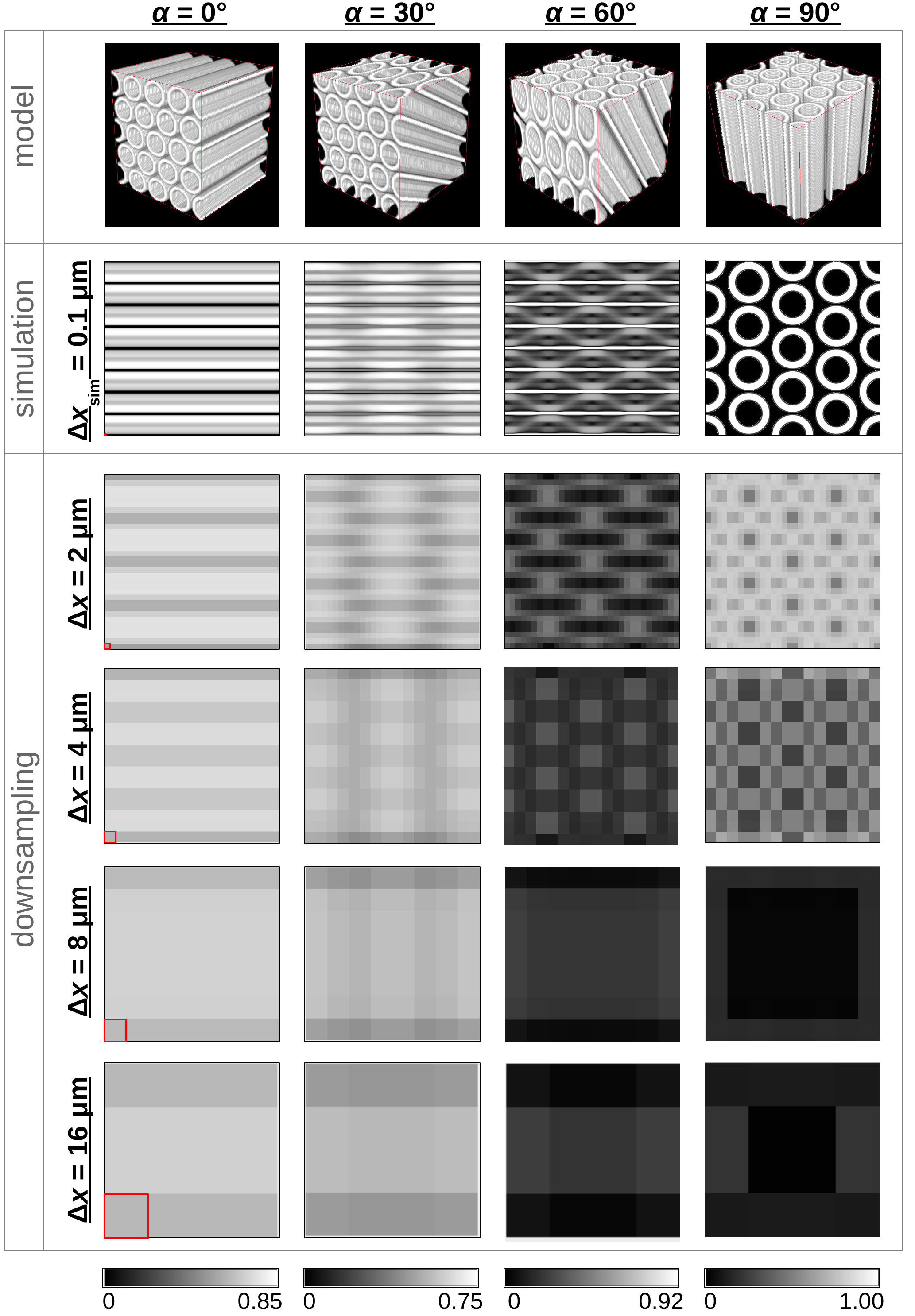}
\caption{Retardation images of the hexagonal fibre bundle for selected fibre inclination angles $\alpha$, simulated according to the microscopic model. The retardation values have been computed from (downsampled) image series with different pixel sizes $\Delta x$ (according to Tab.\ \ref{tab:downsampling}) after the myelin density correction. The pixel sizes are indicated by a square in the left bottom corner of the retardation images. To enhance the image contrast, a different scale bar is used for each inclination angle (the minimum value of the high-resolution retardation image is encoded in black, the maximum value in white).}
\label{fig:fibrebundle_retimage}
\end{figure}


\end{document}